\documentclass[twocolumn, prl, showpacs,superscriptaddress,floatfix]{revtex4}
\usepackage{graphicx}
\usepackage{dcolumn}
\usepackage{bm}
\usepackage{amsmath}

\renewcommand {\vec} {\mathbf}

\begin{document}

\title{Exact solution for infinitely strongly interacting Fermi gases in tight waveguides}
\author{Liming Guan}
\affiliation {Institute of Physics, Chinese Academy of Sciences,
Beijing 100080, China}
\author{Shu Chen}
\email{schen@aphy.iphy.ac.cn}
\affiliation {Institute of Physics,
Chinese Academy of Sciences, Beijing 100080, China}
\author{Yupeng Wang}
\affiliation {Institute of Physics, Chinese Academy of Sciences,
Beijing 100080, China}
\author{Zhong-Qi Ma}
\affiliation {Institute of High Energy Physics, Chinese Academy of
Sciences, Beijing 100049, China}

\begin{abstract}
We present an exact analytical solution of the fundamental systems
of quasi-one-dimensional spin-1/2 fermions with infinite repulsion
for arbitrary confining potential. The eigenfunctions are
constructed by the combination of Girardeau's hard-core contacting
boundary condition and group theoretical method which guarantees
the obtained states to be simultaneously the eigenstates of $S$
and $S_z$ and fulfill the antisymmetry under odd permutation. We
show that the total ground-state density profile behaves like the
polarized noninteracting fermions, whereas the spin-dependent
densities display different properties for different spin
configurations. We also discuss the splitting of the ground states
for large but finite repulsion.
\end{abstract}

\date{\today}
\maketitle


{\it Introduction.---} The experimental progress in manipulating
cold atoms in effective one-dimensional (1D) waveguides
\cite{gorlitz,esslinger} has stimulated extensive theoretical and
experimental study of the 1D strongly correlated atomic systems.
Particulary, the experimental realization of Tonks-Girardeau (TG)
gases \cite{Paredes,Toshiya} has allowed us to study the
fermionization of Bose gas in the strongly interacting limit. More
recently an interacting 1D Fermi gas with tunable interaction
strengths has also been experimentally realized \cite{Moritz05},
which offers the opportunity of studying the 1D Fermi gases even in
the TG limit. To understand the physical properties of the cold atom
in the strongly interacting limit, exact solutions and some refined
methods capable of dealing with strong correlations are especially
important \cite{Yang,Gaudin,Mattis,Girardeau}. In the infinitely
repulsive limit the many-body state of a TG gas can be constructed
via a Bose-Fermi mapping \cite{Girardeau}. Despite its long history,
the generalization to systems including spin degree of freedom is a
highly non-trivial problem and only recently was tackled
\cite{Girardeau07,Deuretzbacher08}. Nevertheless the construction of
the exact wave function for the fundamental system of {\it
indistinguishable} spin-1/2 Fermi gas in the TG limit is still
lacking despite its great importance \cite{note}. Different from the
Bose system whose ground state (GS) is proved to be a degenerated
ferromagnetic state \cite{Lieb02}, the GS of a Fermi system
generally falls into the state with lowest total spin value $S$
\cite{Mattis}. As we shall clarify later, the mixed symmetry of the
spin function renders the generalization of Bose-Fermi mapping to
the spin-1/2 Fermi system difficult and very challenging. In this
work we present for the first time an analytically exact solution of
quasi-1D Fermi gases with infinite repulsion in trapped potentials.

{\it Model.---} We consider a quasi-1d system with N spin-1/2
fermions tightly confined in an elongated potential trap which is
described by an effective 1D Hamiltonian
\begin{equation}
H=\sum_{i=1}^N\left[ -\frac{\hbar ^2}{2m}\frac{\partial ^2}{\partial x_i^2}%
+V(x_i)\right] +g_{1d}\sum_{i<j}\delta (x_i-x_j), \label{H1}
\end{equation}
with $g_{1d}$ being the effective 1D interaction strength
\cite{Olshanii}. For a harmonic potential, $V(x_i)=\frac 12m\omega
_x^2x_i^2.$ Despite intensive research
\cite{Recati,Gao,Astrakharchik}, there has been rarely rigorous
results on the interacting spin-1/2 fermion systems except the
homogenous Yang-Gaudin model \cite{Yang,Gaudin}. Our exact solution
in the strong coupling limit for the fundamental spin-$1/2$ Fermi
system will provide a firm touchstone for various approximate
methods \cite{Recati,Gao,Astrakharchik} and also deepen our
understanding on the few-body system.

{\it Construction of exact ground-state wavefunction.---} In
general, one can represent the many-body wavefunctions in the
space-spin form as $\Psi \left( x_1,\sigma _1;\cdots ;x_N,\sigma
_N\right) .$ Inspired by the seminal work of Girardeau, the effect
of an infinitely strong interaction can be reduced to a hard-core
boundary condition
\begin{equation}
\Psi \left( x_1,\sigma _1;\cdots ;x_N,\sigma _N\right) \mid
_{x_i=x_j,\sigma _i=-\sigma _j}=0.
\end{equation}
In addition the Pauli exclusion principle enforces the boundary
condition $\Psi \left( x_1,\sigma _1;\cdots ;x_N,\sigma _N\right)
\mid _{x_i=x_j,\sigma _i=\sigma _j}=0$ for two particles with the
same spins. Therefore the general contacting boundary condition for
a TG Fermi gas can be represented as $ \Psi \left( x_i=x_j\right)
=0, $ which is irrelevant to the spin configurations.  Now it is
straightforward to observe that the wave function, which is composed
of Slater determinant of $N=N_{\uparrow }+N_{\downarrow }$
orthonormal orbitals $\phi _1(x),\cdots ,\phi _{N}(x)$ occupied by
either component of Fermions, fulfills the above boundary condition.
Explicitly, we have
\begin{equation}
\psi _A (x_1,\ldots ,x_N)=(N!)^{-\frac 12}det[\phi
_j(x_i)]_{i=1,\ldots ,N}^{j=1,\ldots ,N} \label{psiF}
\end{equation}
with $\phi _j(x_i)$ the eigenstate of the single particle
Hamiltonian $H_i=  -\frac{\hbar ^2}{2m}\frac{\partial ^2}{\partial
x_i^2} +V(x_i)$. So far the spin part of wave function is not
considered yet. Since $H$ is spin independent, $H$ is commutable
with the total spin operator $\hat{\vec{S}}=\sum_i
\hat{\vec{S}}_i$, where $\hat{\vec{S}}_i$ is the spin operator of
the ith particle. This implies that the system possesses a global
SU(2) symmetry and the eigenstates of H are simultaneous
eigenstates of $\hat{\vec{S}}^2$ and $\hat{S}_z$. Thus, only the
eigenstates with the largest eigenvalue $S_{z}=S$ are needed to be
considered and the remaining eigenstates can be calculated from
them by the lowering operator $\hat{S}_{-}$. In addition, the
total wave function of N {\it indistinguishable} fermions has to
be antisymmetric under transposition of any two particles.

According to (\ref{psiF}), the GS corresponds to the fully filled
state with the lowest $N$ orbital occupied and excited states are
generated by occupying higher orbitals. Similar to the spinor boson
case, the GS is highly degenerate in the TG limit due to the
different spin configurations. Among the family of degenerate GSs,
the ferromagnetic spin state with $S_{z}=S=N/2$ is a product of all
spins up which is totally symmetric in permutations. The total wave
function, antisymmetric under transpositions, takes a factorized
form $ \Psi =\psi_A \left( x_1,\cdots ,x_N\right) \chi_{1}(1)\cdots
\chi_{1}(N), $ where $\chi_{1}(i)$ denotes the up-spin and
$\chi_{2}(i)$ the down-spin. For the system with fixed up-spin and
down-spin particles, the ferromagnetic state with $S_{z}=N/2-m$  is
also totally symmetric in permutations and degenerated with the
polarized state, where $m\equiv N_{\downarrow}$ and $n\equiv
N_{\uparrow}=N-m$ are the numbers of particles with down-spin and
up-spin, respectively. So far, only the ferromagnetic state is
constructed. An important issue here is to discuss how the GS
degeneracy in the TG limit is split when $g_{1d}$ is large but
finite, or alternatively, to find the GS which could be a good
approximation of the true wave function when the interaction
strength is very large but not infinite. According to Lieb-Mattis
theorem \cite{Mattis}, for finite interaction strength, the state
with lower $S$ has lower GS energy, therefore the GS for the system
with fixed $n$ and $m$ is the state with $S=S_z=N/2-m$. Intuitively,
the repulsive interaction term will contribute a positive energy to
a state with $S< N/2$, but it does not contribute to a ferromagnetic
state with all spins oriented in the same direction, therefore the
Lieb-Mattis theorem seems counter intuitive. One can understand this
problem by noticing that a ferromagnetic state with $S=N/2$ should
occupy N different orbits due to the Pauli principle, whereas for
the state with lower S, the particles with opposite spins are
allowed to occupy overlapping states and thus lower the energy.

The spin function with $S<N/2$, described by a Young diagram
$[n,m]$, is not totally symmetric. Nevertheless, we can still
represent a wave function formally as a product of $\psi_A$ and
$\psi_S$, where $\psi_S$ denotes a symmetric function composed of
linear combination of product of sign functions and spin functions.
Next we shall resort to the group theoretical method to construct
$\psi_S$.

Before presenting our result, we first introduce some notations of
group theory \cite{Ma}. Let $B_{\alpha}=\{b_{1},b_{2}, \ldots ,
b_{m}\}$ be a set of $m$ different integers where $1\leq
b_1<b_2<\ldots <b_m\leq N$. The $n=N-m$ remaining different integers
$a_{1}$, $a_{2}$, $\ldots$, $a_{n}$, satisfying $a_{i}\neq b_{j}$
and $1\leq a_{1}<a_{2}<\ldots <a_{n}\leq N$ are also determined by
the set $B_{\alpha}$. There are $N!/(m!n!)$ different sets
$B_{\alpha}$. $b_{j}=n+j$ when $\alpha=1$. Corresponding to a set
$B_{\alpha}$, we define a permutation $P_{\alpha}$,
\[
P_{\alpha}=\left(
\begin{array}{cccccccc}
1 & 2 & \ldots & n & n+1 & n+2 & \ldots & N \\
a_1 & a_2 & \ldots & a_n & b_1 & b_2 & \ldots & b_m
\end{array}\right) .
\]
Remind that $P_{1}$ is the identical permutation. The left coset of
a subgroup $S_{n}\otimes S_{m}$ of $S_{N}$, where $S_{n}$ and
$S_{m}$ are respectively the permutation groups of the first $n$
objects and the last $m$ objects, is denoted by
$P_{\alpha}\left(S_{n}\otimes S_{m}\right)$. Introduce
$Q_{1}=\prod_{i=1}^n\prod_{j=n+1}^N sgn(x_{i} - x_{j})$, with the
sign function $sgn(x_i- x_j)=(x_i-x_j)/|x_i-x_j|$, and
$Q_{\alpha}=P_{\alpha}Q_{1}$. A spin state with $S=S_{z}=N/2-m$ is
denoted by
$P_{\alpha}\mathcal{Y}_1^{[n,m]}Z_{1}=\mathcal{Y}_{\alpha}^{[n,m]}Z_{\alpha}$,
which is a basis tensor of the tensor subspace
$\mathcal{Y}_{\alpha}^{[n,m]}\mathcal{T}$ of $SU(2)$ of rank $N$
with the highest weight \cite{Ma}, where $Z_{1}=\chi _{1}(1)\ldots
\chi _{1}(n)\chi_{2}(n+1) \ldots \chi_{2}(N)$,
$Z_{\alpha}=P_{\alpha}Z_{1}$,
$\mathcal{Y}_{\alpha}^{[n,m]}=P_{\alpha}\mathcal{Y}_1^{[n,m]}
P_{\alpha}^{-1}$, and $\mathcal{Y}_1^{[n,m]}=\left(\sum_{R\in
S_{n}}R\right)\left(\sum_{T\in S_{m}}T\right)
\left\{\prod_{j=1}^{m}\left[E-(j~n+j)\right]\right\}$, where $E$ is
the identical permutation and $(j~n+j)$ is the transposition between
$j$ and $n+j$. Our definition for a Young operator \cite{Ma}
coincides with that in \cite{boe}, but different from that in
\cite{ham}. The spin states with $S=S_{z}=N/2-m$ based on one
definition are the linear combinations of those on the other.

\textbf{Theorem.~} The totally symmetric wave function constructed
by the product of the sign functions $Q_{\alpha}$ and the basis
tensors $Z_{\alpha}$ is
\begin{eqnarray}
\psi_S &=& \left\{\displaystyle \sum_{\alpha=1}^{N!/(n!m!)}
P_{\alpha}\right\} \left\{
Q_{1}\left(\mathcal{Y}_{1}^{[n,m]}Z_{1}\right) \right\}\nonumber \\
&=& \displaystyle \sum_{\alpha=1}^{N!/(n!m!)}
\left\{\mathcal{Y}_{\alpha}^{[n,m]} Q_{\alpha}\right\} Z_{\alpha}.
\label{psiS}
\end{eqnarray}

\textbf{Proof.} Since $Q_{1}$ and $\mathcal{Y}_{1}^{[n,m]}Z_{1}$
both are invariant in left-multiplying by any element of the
subgroup $S_{n}\otimes S_{m}$, the action of $\sum_{\alpha}
P_{\alpha}$ is proportional to that of the sum over all elements in
$S_{N}$ so that $\psi_{S}$ is invariant in $S_{N}$. The last formula
in (\ref{psiS}) is obtained by rearrangement. \hfill $\diamond$

Now it is easy to check that the ground state
\begin{equation}
\Psi = \psi_A \psi_S
\end{equation}
with $\psi_A$ and $\psi_S$ given by (\ref{psiF}) and (\ref{psiS})
fulfills all the requirements of symmetry and hard-core boundary
condition and are simultaneously the eigenstates of $\hat{S}^2$ and
$\hat{S}_z$ with $S=S_z=(N_{\uparrow}-N_{\downarrow})/2$. As a
concrete example, for $N=3$ with $S=1/2$, $\psi_S=
\sum_{\alpha=1}^{3}(3Q_{\alpha}-1) Z_{\alpha}$, where the identity
$\sum_{\alpha=1}^{3} Q_{\alpha}=1$ is used for simplification. We
note that our constructed exact solution has the same spin structure
described by the Young diagram $[n,m]$ as the true wave function
when the strength is very strong but not infinite \cite{Mattis}.
Therefore, our result is expected to interpolate analytically
between the finite-repulsion case and the limit case with infinite
repulsion.

\begin{figure}
\includegraphics[width=3.0in]{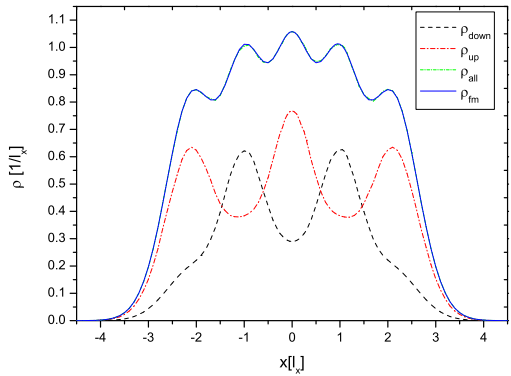}
\caption{ (color online) The GS density distributions of Fermi gas
in the limit of infinite repulsion with $N_{\uparrow}=3$ and
$N_{\downarrow}=2$.} \label{fig1}
\end{figure}

{\it Density distributions.---} The spin-dependent reduced one-body
density matrices are defined as $ \rho_{\sigma} (x,x^{\prime }) = C
\int \prod_{i=2}^{N} dx_{i} \Psi ^{\dagger}\left( x,X'\right)
P_{\sigma}^{(1)} \Psi \left( x^{\prime }, X' \right) , $
where $X'=(x_{2},\cdots,x_{N})$,
$P_{\uparrow,\downarrow}^{(1)}=(1\pm \hat{\sigma}_z^{(1)})/2$ with
$\sigma=\uparrow$ ($\downarrow$) corresponding to $+$ ($-$) and $C$
are normalized constants fixed by the conditions $\int dx
\rho_{\sigma} (x) = N_{\sigma}$. Here the spin-dependent single
particle densities $\rho_{\sigma} (x) =\rho_{\sigma} (x,x)$ are the
diagonal elements of the corresponding reduced density matrices. The
total density is defined as $\rho (x)= \rho_{\uparrow} (x) +
\rho_{\downarrow} (x)$. After some algebra, we can prove that the
total GS density is identical to the density of a polarized
N-particle free fermion system which takes the following simple form
$\rho (x)= \sum_{l} |\phi_{l}(x)|^2$, where the summation is over
the lowest $N=n+m$ single particle states. For the 1D harmonic trap,
the orbital are the oscillator eigenstates. In contrast with the
ferromagnetic ground state where $\rho_{\sigma} (x)= (N_{\sigma}/N)
\rho(x)$, there is no a simple expression of the spin-dependent
density for the general state corresponding to Young diagram
$[n,m]$. Nevertheless, one can calculate $\rho_{\sigma} (x)$
directly from the exact ground state wavefunctions. In Fig.1, we
display the GS density distributions of a five-particle systems
composed of three spin-up and two spin-down fermions. Despite the
same total density distributions, the spin-dependent distributions
are apparently different from that of the ferromagnetic state. Here
the total density profile exhibits five peaks corresponding to that
the wavefunction is composed of the lowest five orbitals. However,
due to the spin-dependent term of eq. (\ref{psiS}), the
spin-dependent density profiles are reorganized so that the spin-up
part and spin-down part avoid overlapping together and show
alternative peak structure to lower the energy. Although the exact
many-body wave function is constructed, the calculation of the
density distribution and momentum distribution for a large system
remains a difficult task due to the time consuming to calculate
multidimensional integrals. Nevertheless, some robust features are
found to be not sensitive to the size. For example, the peaks in the
spin-up and spin-down density distributions appear alternatively and
the density distributions also show the parity symmetry.
\begin{figure}
\includegraphics[width=3.0in]{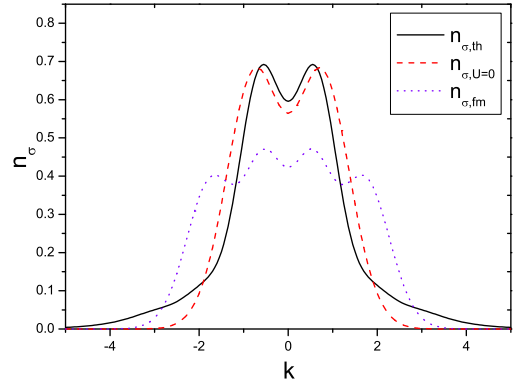}
\caption{ (color online) The GS momentum distributions
$n_{\sigma}(k)$ for the system with $N_{\uparrow}=N_{\downarrow}=2$.
Solid line corresponds to spin singlet, dotted line to ferromagnetic
state and dashed line to the case of free fermion for comparison.}
\label{fig2}
\end{figure}

The momentum distribution can be directly obtained from the Fourier
transformation of the reduced density matrices
$n_{\sigma}(k)=(2\pi)^{-1}\int dx dx' e^{ik(x-x')}\rho_{\sigma}
(x,x')$. For the ferromagnetic ground state,  we have $n_{\sigma}
(k)= (N_{\sigma}/N) \sum_{l=0}^{N-1}|\widetilde{\phi}_{l}(k)|^2$
with $\widetilde{\phi}_{l}(k)$ the Fourier transformation of the
$l$-th oscillator eigenstate. Similarly, such a simple expression
does not generally hold true for the spin state with the Young
diagram $[n,m]$. As shown in Fig.2, the momentum distribution
displays quite different behavior for the spin singlet and
ferromagnetic state. Comparing with the noninteracting case of
$g_{1d}=0$ where $n_{\sigma} (k)=
\sum_{l=0}^{N_{\sigma}-1}|\widetilde{\phi}_{l}(k)|^2$ , the momentum
distribution develops a wide tail.

\begin{figure}
\includegraphics[width=3.7in]{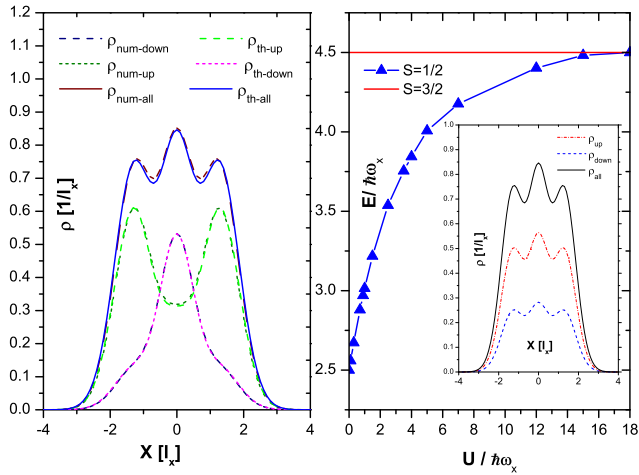}
\caption{ (color online) (a) Comparison of GS density distributions
obtained by ED for the system with $N_{\uparrow}=2$ and
$N_{\downarrow}=1$ with the analytical result of $S=1/2$ state. (b)
The energy versus the interaction for different spin state. The
density distribution for the ferromagnetic state (inset). }
\label{fig3}
\end{figure}

{\it Comparison with system with large but finite repulsion.---}
Apart from infinite repulsion limit, there is no analytical solution
available for the harmonic potential. However, for the small
particle system, we can apply the exact diagonalization method
\cite{Deuretzbacher} to calculate its GS properties and compare with
our analytical result in the infinite limit. In terms of the
fermionic creation and destruction operators $a_{i\sigma }^{\dagger
}$ and $a_{i\sigma }$ of the axial harmonic oscillator, we get the
many-body Hamiltonian corresponding to (\ref{H1})
\begin{equation}
H=\hbar \omega _x\sum_{i,\sigma }(i+\frac 12)\hat{a}_{i\sigma }^{\dagger }%
\hat{a}_{i\sigma }+ U \sum_{ijkl}I_{ijkl}\hat{a}_{i\uparrow
}^{\dagger }\hat{a}_{j\downarrow }^{\dagger }\hat{a}_{k\downarrow }\hat{a}%
_{l\uparrow },  \label{H2}
\end{equation}
where $ I_{ijkl}=l_x\int_{-\infty }^\infty dx\phi _i(x)\phi
_j(x)\phi _k(x)\phi _l(x) $
are the dimensionless interaction integrals with $l_x=\sqrt{\hbar /m\omega _x%
}$ and $U=g_{1d}/l_x$. The Hamiltonian (\ref{H2}) can be exactly
diagonalized in the truncated basis of eigenstates of the harmonic
oscillator and then the GS density can be calculated numerically.
Fig. \ref{fig3}a shows the 3-particle state corresponding to the
Young diagram $[2,1]$, compared to the results obtained from exact
diagonalization (ED) of (\ref{H2}) with parameter $U=15 \hbar
\omega_x$. The numerical results are in perfect agreement with our
exact results, which indicates that the limit of infinite repulsion
is practically reached at $U=15 \hbar \omega_x$
\cite{Deuretzbacher,Hao06}. Instead, as displayed in the inset of
Fig. \ref{fig3}b, the spin-dependent distribution for the degenerate
ferromagnetic GS state displays quite different behaviors from the
true GS of the system with large finite repulsion. We also show how
the GS energy of the state with $S=1/2$ changes versus the increase
of interaction strength in Fig. \ref{fig3}b. In the large repulsion
limit, it tends to become degenerate with the ferromagnetic state.
It would be interesting to compare our results with Ref. \cite{del
Campo}, where the stability of Fermi gases with the presence of
attractive p-wave interaction has been discussed. In the limit case
with absence of p-wave interaction, its phase diagram is consistent
with our conclusion, i.e., the GS is an antiferromagnetic state.

{\it Experimental realizability and detections.---}  As the Feshbach
resonances in qausi-1D 2-component Fermi system have been observed
\cite{Moritz05}, in principle the 2-component fermionic TG gas can
be realized. For the high-dimensional system, the physics in the
infinitely interaction limit (the unitary limit) as well as the BEC
and BCS crossover around the unitary limit have been studied
extensively. The main challenge for realization of quasi-1D Fermi
system comes from that both the Fermi energy and temperature should
be much lower than the transverse confining energy. The
spin-dependent density distribution might be detected within the
current techniques which allows for measurement of specie-dependent
properties\cite{Zwierlein}.

{\it Summary.---} We have constructed the exact eigenstates of the
fundamental system of quasi-1D spin-1/2 fermions with infinite
$\delta$ repulsion by means of group theoretical method. While the
infinite repulsion is described by a hard-core boundary condition,
the group theoretical construction guarantees our wave function
automatically fulfilling the permutation symmetry and being the
eigenstates of $S$ and $S_z$. The construction scheme and the
formula for spin densities are valid independent of the trapping
potential and the particle number. For large but finite repulsion we
have calculated the ground state for a few-particle system
numerically by using the exact diagonalization method. The numerical
result is found to be in excellent agreement with our analytical
result. Our construction of exact eigenstates is valid even when a
Zeeman term in the Hamiltonian exists because it does not change the
nature of the states.

\begin{acknowledgments}
This work is supported by NSF of China Nos. 10821403, 10574150,
10675050 and National Program for Basic Research of MOST, China.
\end{acknowledgments}

\end{document}